\documentclass[lettersize,journal]{IEEEtran}
\usepackage{amsmath,amsfonts}
\usepackage{algorithmic}
\usepackage{algorithm}
\usepackage{array}
\usepackage[caption=false,font=normalsize,labelfont=sf,textfont=sf]{subfig}
\usepackage{textcomp}
\usepackage{stfloats}
\usepackage{url}
\usepackage{verbatim}
\usepackage{graphicx}
\usepackage{cite}
\usepackage{xcolor}
\usepackage{balance}

\definecolor{visioPurple}{RGB}{112,48,160}   
\definecolor{visioLightBlue}{RGB}{0,176,240} 

\begin{document}

\title{A Methodology for Impedance-based Stability Margin Analysis for Interconnected Offshore Wind Clusters}

\author{Germano Rugendo Mugambi,~\IEEEmembership{Student Member,~IEEE,}
        Behnam Nouri,~\IEEEmembership{Member,~IEEE,}
        Oscar Sabor\'{i}o-Romano,~\IEEEmembership{Senior~Member,~IEEE,}
        George Alin Raducu, and Nicolaos A. Cutululis,~\IEEEmembership{Senior~Member,~IEEE}

\thanks{Germano R. Mugambi, Oscar Sabor\'{i}o-Romano, and Nicolaos A. Cutululis are with the Department of Wind and Energy Systems, Technical University of Denmark, Roskilde 4000, Denmark (email: gemuga@dtu.dk; osro@dtu.dk; niac@dtu.dk)..}
\thanks{Germano R. Mugambi and George Alin Raducu are with Vattenfall Vindkraft A/S, Kolding 6000, Denmark (email: germano.mugambi@vattenfall.com; alingeorge.raducu@vattenfall.com).}
\thanks{Behnam Nouri is with Vattenfall Windkraft Europe GmbH, Hamburg 20457,
Germany (email: behnam.nouri@vattenfall.de).}}



\maketitle

\begin{abstract}
With recent developments in offshore grid architectures, power park modules (PPMs) such as clusters of offshore wind power plants (OWPPs) are increasingly interconnected offshore. Consequently, it is necessary to assess how integrating a new OWPP affects the stability margins of an existing OWPP at the point of connection. Although impedance-based methods are widely used for small-signal stability assessment of interconnected converter-based systems, many studies rely primarily on Nyquist encirclements and do not explicitly quantify stability margins. Thus, this paper proposes a general impedance-based methodology to (i) evaluate the stability margins of an existing connection after a new PPM is integrated and (ii) derive a maximum allowable impedance for the new connection such that the minimum stability margin requirements specified by system operators are satisfied and stable operation is maintained. In addition, new Nyquist-based stability regions are introduced to complement the generalized Nyquist criterion, providing analytical indications of margin compliance and headroom. The proposed method is validated through case studies using vendor-based frequency-domain models of two interconnected OWPPs and HVDC system.
\end{abstract}


\section{Introduction}
\IEEEPARstart{R}{ecent} developments in offshore wind cluster design involve configurations in which the transmission system operators (TSOs) own the offshore transmission infrastructure, and offshore wind power plants (OWPPs) connect directly to the switchgear at the HVDC converter station. This approach is already being deployed at the Ijmuiden Ver wind farm by the TSO TenneT \cite{Mugambi2025}. A key stability concern for such topologies is the potential for adverse interactions arising from the converter control systems \cite{interaction}. Furthermore, it is essential to assess how a new connection affects the stability margins of existing OWPP, which must continue to maintain their stable operation and comply with TSO requirements specified at the Point of Connection (PoC). Among these requirements are limits on phase margins at resonance frequencies \cite{LatinovicMilica2025}. Impedance-based analysis has emerged as an effective method for addressing small-signal stability challenges in these interconnected converter-based systems. The approach has previously been applied to identify the root cause of harmonic instability in an HVDC connected OWPP \cite{Buchhagen2016}.

In the impedance-based approach for grid-connected converters, the overall system is partitioned into source and load subsystems, represented by their small-signal impedances/matrices $Z_s(s)$ and $Z_l(s)$, respectively \cite{Sun2011}. The interaction between these two subsystems can be represented as a closed-loop feedback system characterized by the loop gain matrix for multi‑input multi‑output (MIMO) systems or impedance ratio for single-input single-output (SISO) case $L(s) = Z_s(s)Z_l^{-1}(s)$. The system stability is evaluated using the Nyquist stability criterion (NSC) by checking whether the Nyquist plot of $L(s)$ encircles the point $(-1,0j)$; encirclement indicates instability, while its absence implies stability \cite{AminSSS,Francisco24}. The application of the NSC assumes that each subsystem is individually stable, i.e., it contains no right-half-plane (RHP) poles. Due to its maturity, well‑defined stability metrics, and compatibility with black‑box converter models, the impedance-based method is increasingly being adopted within industry for small‑signal stability assessments. Notably, TenneT and other European TSOs are requesting impedance-based studies over a wide frequency range (e.g., 10~Hz--2500~Hz), alongside EMT-based studies \cite{Behnam_Nouri2025,LatinovicMilica2025}.

Although now widely applied to grid-connected converters, the impedance-based approach originates from earlier work on dc–dc power converters. Middlebrook first introduced the method in 1976, demonstrating that system stability could be ensured by keeping the minor loop gain defined as the ratio of the filter impedance to the converter input impedance within the unit circle of the complex plane \cite{Middlebrook}. Subsequently, this concept was extended to distributed power systems, where the minor loop gain was required to avoid a predefined forbidden region in the complex plane \cite{Wildrick93,Wildrick95}. 

\IEEEpubidadjcol
Building on this foundation, further work introduced new impedance specifications for load impedance and developed updated forbidden regions\cite{Feng99,Feng202}. The resulting design constraint required the load impedance to remain within a prescribed range to ensure small-signal stability and adequate decoupling between subsystems. However, because the forbidden region is based on a fixed phase margin, the criterion implicitly assumes that any new load that satisfies the stability margins with the fixed source will not affect the overall system stability. This is a conservative design, as newly introduced impedances can create additional resonance frequencies that impact the margins of existing loads. Thus, the stability margin criteria defined do not dynamically update when new loads are added, limiting the method's applicability in converter-based systems.

In \cite{Wang2020}, it is proposed to specify impedance profiles even when there are open-loop right-half-plane (RHP) poles by determining the number of open-loop RHP poles and the number of encirclements. While the approach enables stability assessment in the presence of open‑loop RHP poles; it only determines stability through Nyquist encirclements and therefore provides no quantitative measure of stability margin, and secondly, it does not evaluate how the integration of a new converter system impacts the existing converter stability margins. Furthermore, numerical methods are required to determine the number of RHP in the loop gain, making the approach highly undesirable in practice \cite{Sun22}. Moreover, the test system consists of only two converters operating in parallel, which does not reflect a real OWPP scenario composed of multiple strings with numerous converters and an inter-array cable network.

This paper proposes a generalized impedance-based stability margin analysis method for interconnected power park module (PPM) clusters. The method enables impedance specification for a newly connected PPM such that the existing PPM maintains the prescribed stability margins and consequently stable interoperability. The approach is validated using vendor-based EMT project models and frequency-domain models. In addition, the method quantifies the impact of the new connection on the stability margins of the existing PPM. Furthermore, the paper introduces new Nyquist-based stability regions that replace conservative forbidden regions and enable the assessment of margin compliance and headroom. Unlike prior impedance-based criteria that only ensure no Nyquist encirclement or use fixed forbidden regions, the methodology quantifies the change in stability margins when adding a new PPM and provides a condition to guarantee compliance and stable operation.

\section{Impedance modeling}
\label{impedance modeling}
\begin{figure}
    \centering
    \includegraphics[width=1\linewidth]{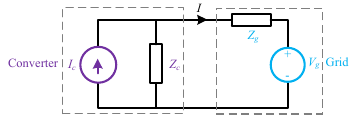}
    \caption{Small-signal representation of grid-connected converter \cite{Sun2011}.}
    \label{fig:con-grid}
\end{figure}

In the impedance-based stability criterion, the impedance response of a grid-connected converter is obtained by partitioning the overall system into two subsystems, as shown in Fig. \ref{fig:con-grid}. The grid-connected converter is modeled as a Norton equivalent with an internal impedance $Z_{c}(s)$, while the grid is represented by a Thévenin equivalent with impedance $Z_{g}(s)$. The terminal impedances  $Z_{c}(s)$ and $Z_{g}(s)$ are then used to determine small-signal dynamics. Applying Kirchhoff's circuit laws to the interconnected system in Fig. \ref{fig:con-grid}, a closed-loop feedback representation is obtained, which can be used to determine the system stability \cite{Cheah-Mane2023}. Based on Fig. \ref{fig:con-grid}, the converter output current is 
\begin{equation}
I(s)
= 
\frac{Z_{c}(s)\, I_{c}(s) - V_{g}(s)}
{Z_{c}(s) + Z_{g}(s)} .
\label{eq:eq1}
\end{equation}

Equation \ref{eq:eq1} can be rearranged to
\begin{equation}
I(s)
=
\left[
I_{c}(s)
-
\frac{V_{g}(s)}{Z_{c}(s)}
\right]
\cdot
\frac{1}{1 + \textcolor{visioLightBlue}{Z_{g}(s)}/\textcolor{visioPurple}{Z_{c}(s)}} .
\label{eq:eq2}
\end{equation}

The ratio $Z_g(s)/Z_c(s)$ constitutes the minor-loop gain $L(s)$ of the negative feedback loop. The system stability can be assessed using the Nyquist plot of $L(s)$ based on the NSC. For the Nyquist criterion to be applicable, the grid is assumed to be stable without the converter, and the converter is assumed to be stable when connected to an ideal grid. Under these assumptions, the closed-loop stability can be determined from the loop-gain Nyquist plot without checking for possible RHP poles \cite{Sun22}.

When the frequency coupling effect between the converter and grid is considered, the impedances $Z_g(s)$ and $Z_c(s)$ are represented as 2x2 matrices yielding a MIMO feedback system \cite{ShahMIMO,Nouri2021}. In this case, stability is assessed using the determinant or eigenvalue decomposition of $I + L(s)$ \cite{Maciejowski89,Z-toolFrancisco}. A determinant‑based approach is often preferable because it requires only a single Nyquist plot. In contrast, the eigenvalue-based approach requires tracking multiple eigenvalue trajectories across frequency, which can become ambiguous in larger systems. Nonetheless, eigenvalue analysis remains useful when individual eigenvalues can be linked to specific physical behaviors of the system \cite{Sun22}. The coupling between the control functions is tight for frequencies below 100~Hz and can introduce negative damping, especially near the fundamental frequency. However, above 100~Hz the coupling is weaker and can be neglected in sub-synchronous oscillation (SSO) studies \cite{sunTwoPort_1}.

The standard approach for the impedance-based method faces scalability issues when applied to a large system with multiple converters. This is because $Z_g(s)$ and $Z_c(s)$ grow in dimension with the number of connected converters, making impedance extraction cumbersome. Moreover, obtaining the network impedance matrix $Z_g(s)$ for a practical power system requires numerous frequency-scan experiments to ensure accuracy, which is tedious and time-consuming \cite{AEMO_Report}. One alternative is to aggregate individually stable converters or inverter-based resources (IBRs) into the classical two equivalent subsystems. However, such encapsulation does not necessarily guarantee the absence of RHP poles, particularly when HVDC converters and multi-terminal grids are involved \cite{Francisco24}. Still, if performed systematically over all relevant combinations and partitions, the SISO formulation introduced in \cite{Sun2011} can be sufficient \cite{Sun22}. 

Another option is to use harmonic linearization to model component impedances. In this method, components are represented in the sequence domain (positive and negative), and under balanced conditions, there is no coupling between the two sequences \cite{Sun2009,sunTwoPort_1,Sun22}. Each component (or converter) is then replaced by its sequence impedance, reducing the network to a linear impedance network whose stability can be assessed using frequency-domain techniques such as impedance-based analysis \cite{Sun22}. This method is widely adopted in industry practice for harmonic stability assessment. The stability can then be evaluated independently for each sequence.

\begin{figure}
    \centering
    \includegraphics[width=1\linewidth]{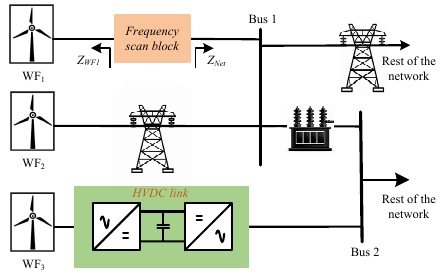}
    \caption{Wide area network impedance scan setup.}
    \label{fig: wide_area}
\end{figure}

A further alternative is the reverse impedance-based approach, which deviates from the standard formulation by assuming the grid remains stable when the IBR is connected. Stability following disconnection can then be assessed using the Nyquist plot of the loop gain. In this framework, stability is evaluated by focusing on one selected IBR at a time, while the remainder of the grid (including other IBRs) is represented by an equivalent impedance/admittance as seen from the selected IBR’s point of connection. A wide-area network scan is performed in an EMT model of the full system at the selected IBR to obtain both the IBR impedance and the grid impedance as viewed from that connection point \cite{NREL}. This approach has been applied to the West Murray Zone power system in Australia to analyze SSO driven by solar farms \cite{AEMO_Report}. Fig. \ref{fig: wide_area} illustrates the wide-area network scan configuration with focus on wind farm 1.

Based on the analysis above, this study employs harmonic linearization and the reverse approach to obtain positive- and negative-sequence impedances, which are then used to define impedance specifications and assess stability margins of the existing string (or PPM) when a parallel string (or PPM) is connected. Harmonic linearization is employed to derive a sequence-domain model that decouples the positive- and negative-sequence components under balanced operating conditions and can be systematically extended to multi-converter systems, enabling standard impedance-based stability analysis. The reverse approach is used to match practical operation (stable with IBRs connected) and to extract the grid impedance seen at the point of connection via wide-area network scans without forming a high-order network impedance matrix.

\section{OWPP Impedance Formulation}
\label{Z_formulation}
\begin{figure}
    \centering
    \includegraphics[width=1\linewidth]{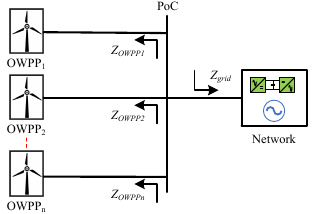}
    \caption{Offshore wind power plant cluster configuration.}
    \label{fig:Offshore_config}
\end{figure}

When comparing offshore wind farm clusters, the offshore grid represented by the Network in Fig. \ref{fig:Offshore_config} to which they are connected can be treated as sufficiently certain, since the cable network and converter models do not change. Hence, for a given operating point, the offshore grid can be assumed fixed. Three alternative representations are considered: (i) a Thévenin-equivalent grid whose short-circuit power is set to match the total short-circuit strength of the full HVDC link, (ii) an explicit full HVDC link model, and (iii) a frequency-dependent representation based on the HVDC link harmonic impedances. 

The individual impedances in Fig.~\ref{fig:Offshore_config} are
$Z_{\mathrm{OWPP}1}, Z_{\mathrm{OWPP}2}, \dots, Z_{\mathrm{OWPP}n}, Z_{\mathrm{grid}}$.
Then, the network impedance seen by $\mathrm{OWPP}1$ is
\begin{equation}
Z_{\mathrm{net,old}}
= Z_{\parallel}\!\bigl(Z_{\mathrm{grid}}, Z_{\mathrm{OWPP}2}, \dots, Z_{\mathrm{OWPP}n}\bigr).
\end{equation}
where $Z_{\mathrm{net,old}}$ denotes the equivalent impedance of the network excluding $\mathrm{OWPP}1$ (i.e., the grid in parallel with the remaining OWPPs) prior to connecting the new wind farm. Based on the minor-loop definition in \eqref{eq:eq2}, the loop gain is defined as
\begin{equation}
\label{eq:L_old}
L_{\mathrm{old}}(j\omega)
= \frac{Z_{\mathrm{net,old}}(j\omega)}{Z_{\mathrm{OWPP}1}(j\omega)}.
\end{equation}

Let $Z_{\mathrm{OWPP}(n+1)}$ denote the impedance of the newly added wind farm. Since the new farm is connected in parallel at the PCC with the existing network, the updated network impedance is given by
\begin{equation}
\label{eq:Znet_new}
Z_{\mathrm{net,new}}
= Z_{\mathrm{net,old}} \parallel Z_{\mathrm{OWPP}(n+1)}
= \frac{Z_{\mathrm{net,old}}\,Z_{\mathrm{OWPP}(n+1)}}{Z_{\mathrm{net,old}}+Z_{\mathrm{OWPP}(n+1)}}.
\end{equation}
Accordingly, the new minor-loop gain becomes
\begin{align}
\label{eq:Lnew_expand}
L_{\mathrm{new}}(j\omega)
&= \frac{Z_{\mathrm{net,new}}(j\omega)}{Z_{\mathrm{OWPP}1}(j\omega)} \nonumber\\
&= \frac{Z_{\mathrm{net,old}}}{Z_{\mathrm{OWPP}1}}\,
\frac{Z_{\mathrm{OWPP}(n+1)}}{Z_{\mathrm{net,old}}+Z_{\mathrm{OWPP}(n+1)}} \nonumber\\
&= L_{\mathrm{old}}(j\omega)\;
\frac{1}{1+\dfrac{Z_{\mathrm{net,old}}(j\omega)}{Z_{\mathrm{OWPP}(n+1)}(j\omega)}}.
\end{align}

Define the impedance ratio as
\begin{equation}
\label{eq:rho_def}
\rho(j\omega) = \frac{Z_{\mathrm{net,old}}(j\omega)}{Z_{\mathrm{OWPP}(n+1)}(j\omega)}.
\end{equation}

Hence, the new minor-loop gain is
\begin{equation}
\label{eq:Lnew_rho}
L_{\mathrm{new}}(j\omega)=\frac{L_{\mathrm{old}}(j\omega)}{1+\rho(j\omega)}.
\end{equation}

\subsection{Phase and Gain Margins Calculation}

The phase margin (PM) is evaluated at the gain-crossover ($gc$) frequency, defined by
$\lvert L_{\mathrm{new}}(j\omega_{gc})\rvert = 1$, i.e., where the magnitude of the minor-loop gain is unity. It corresponds to the phase angle of the loop gain at the point where the Nyquist plot crosses the unit circle in the complex plane.
The gain margin (GM) is evaluated at the phase-crossover ($pc$) frequency, defined by
$\angle L_{\mathrm{new}}(j\omega_{pc}) = -180^\circ$, i.e., where the phase of the loop gain crosses $-180^\circ$.

The newly connected OWPP may introduce additional resonance points or shift existing resonance points. Therefore, the stability margins should be evaluated at the updated crossover frequencies as follows.
\begin{align}
\mathrm{PM}_{\mathrm{new}}
&= 180^\circ + \angle L_{\mathrm{new}}(j\omega_{gc,\mathrm{new}}).
\label{eq:PM_new}\\
\mathrm{GM}_{\mathrm{new}}
&= \frac{1}{\lvert L_{\mathrm{new}}(j\omega_{pc,\mathrm{new}})\rvert}. \label{eq:GM_new}
\end{align}
The stability margins can also be expressed in terms of the old loop gain evaluated at the new crossover frequencies. From \eqref{eq:Lnew_rho}, let
\begin{equation}
\label{eq:F_def}
F(j\omega)=\frac{1}{1+\rho(j\omega)}.
\end{equation}
Then,
\begin{equation}
\label{eq:Lnew_old_F}
L_{\mathrm{new}}(j\omega)=L_{\mathrm{old}}(j\omega)\,F(j\omega).
\end{equation}
Accordingly, the phase and gain margins can be written as
\begin{align}
\mathrm{PM}_{\mathrm{new}}
&= 180^\circ + \angle L_{\mathrm{new}}(j\omega_{gc,\mathrm{new}}) \nonumber\\
&= 180^\circ + \angle L_{\mathrm{old}}(j\omega_{gc,\mathrm{new}})
+ \angle F(j\omega_{gc,\mathrm{new}}) \nonumber\\
&= \underbrace{180^\circ + \angle L_{\mathrm{old}}(j\omega_{gc,\mathrm{new}})}_{\mathrm{PM}_{\mathrm{old_{newgc}}}}
- \angle\!\bigl(1+\rho(j\omega_{gc,\mathrm{new}})\bigr), \label{eq:PM_new_from_old}\\[4pt]
\mathrm{GM}_{\mathrm{new}}
&= \frac{1}{\lvert L_{\mathrm{new}}(j\omega_{pc,\mathrm{new}})\rvert} \nonumber\\
&= \frac{1}{\lvert L_{\mathrm{old}}(j\omega_{pc,\mathrm{new}})\rvert\,
\lvert F(j\omega_{pc,\mathrm{new}})\rvert} \nonumber\\
&= \frac{\lvert 1+\rho(j\omega_{pc,\mathrm{new}})\rvert}
{\lvert L_{\mathrm{old}}(j\omega_{pc,\mathrm{new}})\rvert}. \label{eq:GM_new_from_old}
\end{align}

Based on \eqref{eq:Lnew_rho} and \eqref{eq:GM_new_from_old}, the following conclusions can be drawn.

\paragraph{Open-circuit condition of the newly added OWPP}
If $Z_{\mathrm{OWPP}(n+1)} \to \infty$ near the crossover frequencies, then $\rho(j\omega)\to 0$, and thus
$L_{\mathrm{new}}(j\omega)\to L_{\mathrm{old}}(j\omega)$, i.e., the loop gain remains unchanged.

\paragraph{Very low impedance of the newly added OWPP}
If $Z_{\mathrm{OWPP}(n+1)} \to 0$ near the crossover frequencies, then $\rho(j\omega)\to \infty$, and
\begin{equation}
L_{\mathrm{new}}(j\omega)\approx \frac{L_{\mathrm{old}}(j\omega)}{1+\rho(j\omega)} \to 0.
\end{equation}
Therefore, the loop-gain magnitude decreases significantly, which increases the GM and typically increases the PM (i.e., the Nyquist locus is pulled toward the origin).

\subsection{Calculation of $\angle L_{\mathrm{old}}(j\omega)$ at New Crossover Frequencies}

The crossover frequencies appearing in \eqref{eq:PM_new_from_old} can also be interpreted as a set (or range) of critical frequencies specified by the system operator, such that a minimum phase margin must be guaranteed if a resonance point falls within this range. Therefore, it is important to evaluate the new phase margin using the pre-connection loop gain, i.e., $L_{\mathrm{old}}(j\omega)$.

\begin{figure}
    \centering
    \includegraphics[width=1\linewidth]{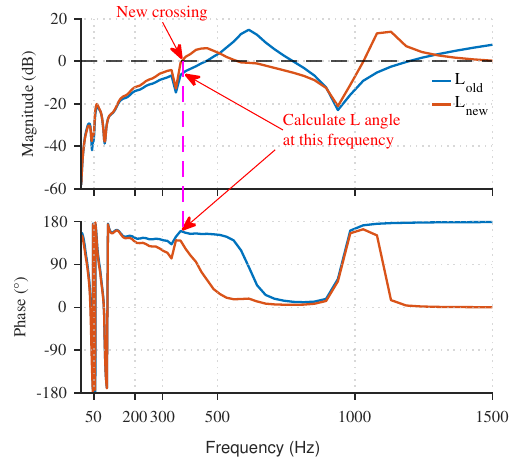}
    \caption{Calculation of $\angle L_{\mathrm{old}}(j\omega)$ at new crossings.}
    \label{fig:L_calculate}
\end{figure}

In particular, the phase of the old loop gain at the new gain-crossover frequency, $\angle L_{\mathrm{old}}(j\omega_{gc,\mathrm{new}})$, can be obtained from the Bode plot of $L_{\mathrm{old}}(j\omega)$, as illustrated in Fig.~\ref{fig:L_calculate}. This calculation is performed automatically from the impedance data, without requiring time-domain simulation or explicit plotting of the Bode diagrams.

\subsection{Impedance Specification}

Recently, some system operators have required that each OWPP maintain minimum stability margins at the PoC at resonance frequencies. Depending on the project, this requirement may apply to each string, which can be treated as an individual power plant. For example, TenneT specifies minimum phase margins of $15^\circ$ and $30^\circ$ for offshore and onshore connections, respectively. The onshore grid typically requires a higher margin because its meshed structure introduces higher levels of uncertainty and a larger number of possible resonance points \cite{LatinovicMilica2025}. In practice, greater emphasis is placed on the phase margin, because a phase crossover is not always guaranteed, whereas a phase margin exists for any gain crossover. To ensure that each newly connected string/OWPP does not reduce the stability margins of the existing plant, a method for specifying the maximum allowable impedance is formulated in this subsection.

Accordingly, based on \eqref{eq:PM_new_from_old}, an inequality can be derived to constrain the impedance of the newly connected OWPP. For brevity, the argument $(j\omega)$ is omitted, $\mathrm{PM}_{\mathrm{old{newgc}}}$ is denoted by $\mathrm{PM}_{\mathrm{old}}$, and $\rho$ is replaced by its definition in \eqref{eq:rho_def}.

Starting from the phase-margin expression,
\begin{equation}
\label{eq:PM_new_start}
\mathrm{PM}_{\mathrm{new}}
= \mathrm{PM}_{\mathrm{old}} - \angle\!\left( 1 + \frac{Z_{\mathrm{net,old}}}{Z_{\mathrm{OWPP,new}}} \right),
\end{equation}
and imposing the minimum phase-margin requirement, $\mathrm{PM}_{\mathrm{new}} \ge \mathrm{PM}_{\min}$, yields
\begin{equation}
\label{eq:PM_angle_bound}
\angle\!\left(1 + \frac{Z_{\mathrm{net,old}}}{Z_{\mathrm{OWPP,new}}}\right)
\le \mathrm{PM}_{\mathrm{old}} - \mathrm{PM}_{\min}.
\end{equation}
Define $\Delta \mathrm{PM} = \mathrm{PM}_{\mathrm{old}} - \mathrm{PM}_{\min}$, and let
\begin{equation}
\label{eq:polar_repr}
1+\frac{Z_{\mathrm{net,old}}}{Z_{\mathrm{OWPP,new}}}=r e^{j\theta},
\end{equation}
where $r>0$ and $\theta=\angle\!\left(1+Z_{\mathrm{net,old}}/Z_{\mathrm{OWPP,new}}\right)$. Then,
\begin{equation}
\label{eq:ratio_mag_re}
\left|\frac{Z_{\mathrm{net,old}}}{Z_{\mathrm{OWPP,new}}}\right|
=|r e^{j\theta}-1|.
\end{equation}
Using the geometric bound $|r e^{j\theta}-1|\ge |e^{j\theta}-1|$, together with the identity $|e^{j\theta}-1|=2|\sin(\theta/2)|$, gives
\begin{equation}
\label{eq:ratio_lower_bound_theta}
\left|\frac{Z_{\mathrm{net,old}}}{Z_{\mathrm{OWPP,new}}}\right|
\ge 2\left|\sin\!\left(\frac{\theta}{2}\right)\right|.
\end{equation}
Since \eqref{eq:PM_angle_bound} implies $\theta \le \Delta \mathrm{PM}$, a sufficient condition is
\begin{equation}
\label{eq:ratio_lower_bound_dPM}
\left|\frac{Z_{\mathrm{net,old}}}{Z_{\mathrm{OWPP,new}}}\right|
\ge 2\sin\!\left(\frac{\Delta \mathrm{PM}}{2}\right).
\end{equation}

Therefore, the impedance of the newly connected OWPP must satisfy
\begin{equation}
\label{eq:Zowpp_new_bound}
|Z_{\mathrm{OWPP,new}}|
\le
\frac{|Z_{\mathrm{net,old}}|}
{2\sin\!\left(\dfrac{\Delta \mathrm{PM}}{2}\right)}.
\end{equation}
Evaluated at the relevant gain-crossover (or critical) frequency $\omega_{gc}$, the design constraint becomes
\begin{equation}
\label{eq:Zowpp_new_bound_wgc}
\left| Z_{\mathrm{OWPP,new}}(j\omega_{gc}) \right|
\le
\frac{\left| Z_{\mathrm{net,old}}(j\omega_{gc}) \right|}
{2 \sin\!\left(
\dfrac{\mathrm{PM}_{\mathrm{old}}(j\omega_{gc})-PM_{\min}}{2}
\right)}.
\end{equation}
Equation \eqref{eq:Zowpp_new_bound_wgc} indicates that at the critical gain-crossover frequency, the new OWPP’s impedance must not exceed a certain fraction of the original network impedance (scaled by a factor related to the difference between the original margin and the minimum required margin). Intuitively, this ensures the new PPM is ‘weak enough’ at that frequency to maintain adequate phase margin.

\subsection{Critical and Caution Areas Concept}

The generalized Nyquist stability criterion assesses closed-loop stability by evaluating the encirclements of the critical point $(-1,0)$. However, encirclement alone provides limited information on robustness, i.e., the available distance to operator-specified minimum phase- and gain-margin requirements. Since system operators prescribe minimum margins for offshore and onshore connections, it is useful to complement the Nyquist plot with a graphical indication of margin compliance and headroom.

Accordingly, this paper augments the Nyquist plot with two \emph{angle-wedge} regions referenced to the negative real axis. The wedges are interpreted at the gain-crossover frequencies (where $\lvert L(j\omega)\rvert=1$), consistent with the definition of phase margin. The \emph{critical area} corresponds to phase margins below the minimum requirement, i.e., $\mathrm{PM}<\mathrm{PM}_{\min}$, indicating noncompliance and reduced stability margins (even if the encirclement condition for stability is still satisfied) which may lead to instability under disturbances. The \emph{caution area} is defined by $\mathrm{PM}_{\min}\le \mathrm{PM}<\mathrm{PM}_{\mathrm{cau}}$, where $\mathrm{PM}_{\mathrm{cau}}$ is a user-selected warning threshold, indicating limited phase-margin headroom.

To also visualize the minimum gain-margin requirement, a circle centered at the origin is included with radius $1/\mathrm{GM}_{\mathrm{lin}}$, where $\mathrm{GM}_{\mathrm{lin}}=10^{\mathrm{GM}_{\mathrm{dB}}/20}$.
At the phase-crossover frequency (i.e., $\angle L(j\omega)=-180^\circ$), satisfying the gain-margin requirement is equivalent to $\lvert L(j\omega)\rvert \le 1/\mathrm{GM}_{\mathrm{lin}}$; hence, an intersection with the negative real axis outside this circle indicates a gain-margin violation.

\begin{figure}
    \centering
    \includegraphics[width=1\linewidth]{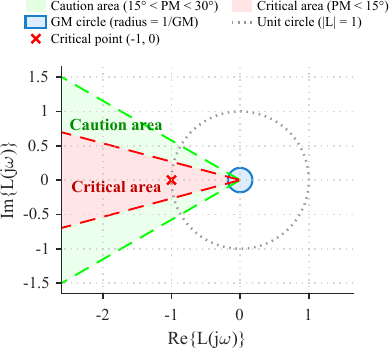}
    \caption{Proposed stability regions.}
    \label{fig:regions}
\end{figure}

To illustrate the proposed regions, $\mathrm{PM}_{\min}=15^\circ$ and $\mathrm{GM}_{\min}=15$~dB (TenneT offshore requirement) are used, and the caution threshold is set to $\mathrm{PM}_{\mathrm{cau}}=30^\circ$ for demonstration. The concept is illustrated in Fig. \ref{fig:regions}.

\subsection{Steps for Impedance Specification and Stability Assessment}
The steps to specify the impedance and assess the stability margins using the proposed method are as follows:
\begin{enumerate}
\renewcommand{\labelenumi}{\arabic{enumi})}
    \item \textit{Step~1:} Perform time-domain EMT simulations for the post-connection configuration to verify that no adverse oscillations are introduced after the new PPM is connected.

    \item \textit{Step~2:} Obtain the PPM and network impedances via frequency scanning (sweep) of the EMT model or from frequency-domain models. Steps~1 and~2 can be omitted if the required impedance data is already available.

    \item \textit{Step~3:} Identify the relevant gain-crossover frequencies, or a set (range) of critical frequencies specified by the system operator. Compute the phase of the pre-connection loop gain, $\angle L_{\mathrm{old}}(j\omega)$, at these frequencies as described in Fig.~\ref{fig:L_calculate}. In addition, evaluate the magnitude of the pre-connection network impedance, $\lvert Z_{\mathrm{net,old}}(j\omega)\rvert$, at the same frequencies.

    \item \textit{Step~4:} Using the first expression in \eqref{eq:PM_new_from_old}, compute the phase margin headroom of the existing connection at the selected frequencies, i.e., $\Delta \mathrm{PM} = \mathrm{PM}_{\mathrm{old}}-\mathrm{PM}_{\min}$.

    \item \textit{Step~5:} Determine the maximum allowable impedance of the new PPM at the selected frequencies using \eqref{eq:Zowpp_new_bound_wgc}. The new connection should satisfy $\lvert Z_{\mathrm{PPM,new}}(j\omega)\rvert \le Z_{\mathrm{limit}}(j\omega)$ to guarantee that the existing PPM maintains the minimum phase-margin requirement.

    \item \textit{Step~6:} Form the updated loop gain using the specified impedance of the new PPM and assess its impact using the Nyquist plot. The Nyquist locus of $L_{\mathrm{new}}(j\omega)$ should not intersect the unit circle within the critical area (red shaded) in Fig.~\ref{fig:regions}; otherwise, the minimum phase-margin requirement is violated at one or more critical frequencies.
\end{enumerate}

\section{Study Model and Study Cases}

\begin{figure}
    \centering
    \includegraphics[width=1\linewidth]{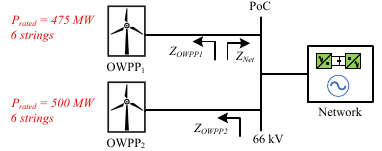}
    \caption{Case~1: Study model configuration with a common transformer.}
    \label{fig:study_case1}
\end{figure}

\begin{figure}
    \centering
    \includegraphics[width=1\linewidth]{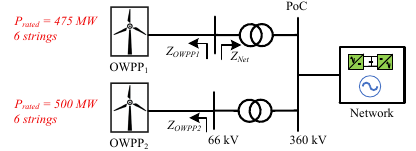}
    \caption{Case~2: Study model configuration with separate transformers.}
    \label{fig:study_case2}
\end{figure}

The conclusions drawn from the analytical formulation in \eqref{eq:Lnew_rho} and \eqref{eq:GM_new_from_old}, together with the design constraint in \eqref{eq:Zowpp_new_bound_wgc}, are validated using the study cases shown in Figs. \ref{fig:study_case1} and \ref{fig:study_case2}. A detailed EMT model is implemented in PSCAD and is based on original equipment manufacturer (OEM)-supplied vendor models from an actual project. Each OWPP comprises six strings; $\mathrm{OWPP}1$ is rated at 475~MW and $\mathrm{OWPP}2$ at 500~MW. For each string, the wind turbines and inter-array cables are aggregated. The turbines operate in grid-following mode with active- and reactive-power control. For the HVDC-connected configuration, the offshore MMC operates in grid-forming mode, whereas the onshore MMC operates in grid-following mode and is connected to a detailed onshore grid model.

Since the analysis is based on the minor-loop gain, the OWPP and network impedances of the EMT models are obtained via frequency scanning. The PSCAD-based scanning tool developed by the Danish TSO Energinet is used \cite{Energinet2025}. The network is represented either by a Thévenin equivalent whose short-circuit strength matches that of the HVDC link, or by the full HVDC link model. In addition, a frequency-domain model is developed in DIgSILENT PowerFactory to obtain sequence-domain impedances, where the harmonic impedances of each component are derived from frequency perturbations and field measurements. The stability-margin assessment is performed at 66~kV for both the Thévenin-equivalent grid and the frequency-domain model, considering configurations with a common transformer and with separate transformers. 


\subsection{Stability Margins Assessment}
The stability-margin assessment is based on the minor-loop gain described in Section~\ref{impedance modeling}. The loop gain $L(j\omega)$ is computed in the sequence domain using the ratio $Z_{\mathrm{net}}/Z_{\mathrm{OWPP}1}$, or equivalently using \eqref{eq:Lnew_rho}. If coupling effects need to be evaluated, the eigenvalues of the loop-gain matrix are computed, i.e., $\lambda\!\left(L(j\omega)\right)$. The impedances used for the analysis are based on reference power set points i.e active power is set to 1 pu while reactive power is set to 0 pu.

First, the analytical loop-gain formulation in \eqref{eq:Lnew_rho} is validated by comparison with the loop gain obtained from the wide-area network scanning procedure in Fig. \ref{fig: wide_area}. In the \emph{calculated} loop gain, the impedances of each OWPP and the grid are obtained separately and then combined according to \eqref{eq:Lnew_rho}. In the \emph{measured} loop gain, the impedances of $\mathrm{OWPP}1$ and the remaining network are obtained directly from the setup in Figs. \ref{fig:study_case1} and \ref{fig:study_case2} via frequency scanning with 1~Hz resolution. The loop gain computed using the proposed formulation matches the measured loop gain, as shown in Fig.~8.
\begin{figure}
    \centering
    \includegraphics[width=1\linewidth]{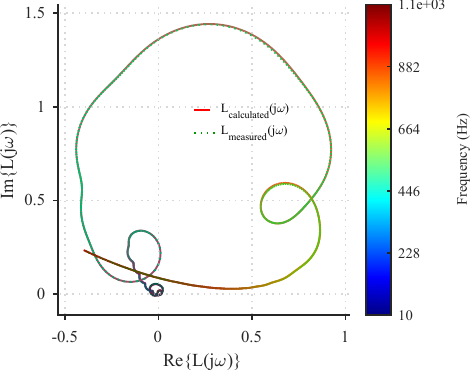}
    \caption{Comparison of calculated and measured loop gain.}
    \label{fig:Loop_compare}
\end{figure}

\begin{figure}
    \centering
    \includegraphics[width=1\linewidth]{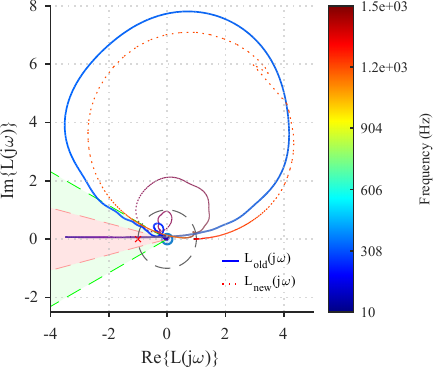}
\caption{Case~1: Stability margins using a Thévenin-equivalent grid and positive-sequence impedance.}
    \label{fig:margins_Thévenin_pos}
\end{figure}

\begin{figure}
    \centering
    \includegraphics[width=1\linewidth]{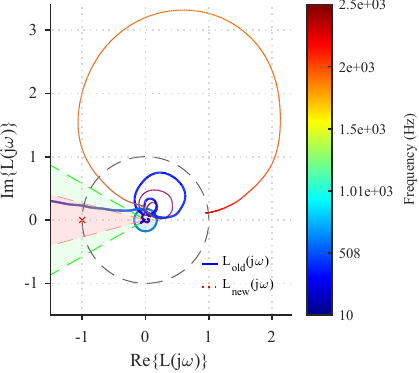}
    \caption{Case~1: Stability margins using frequency-domain model and positive-sequence impedance.}
    \label{fig:margins_harmonic_pos}
\end{figure}

Figs.~\ref{fig:margins_Thévenin_pos} and \ref{fig:margins_harmonic_pos} show the stability margins obtained using (i) a Thévenin-equivalent grid and (ii) the frequency-domain model, respectively. Here, $L_{\mathrm{old}}$ denotes the minor-loop gain of $\mathrm{OWPP}1$ when only $\mathrm{OWPP}1$ is connected, whereas $L_{\mathrm{new}}$ denotes the minor-loop gain of $\mathrm{OWPP}1$ after connecting $\mathrm{OWPP}2$. In both cases, the stability margins of $\mathrm{OWPP}1$ improve after integrating $\mathrm{OWPP}2$, which corroborates the conclusion in Section~\ref{Z_formulation} derived from \eqref{eq:Lnew_rho}. 

\begin{figure}
    \centering
    \includegraphics[width=0.95\linewidth]{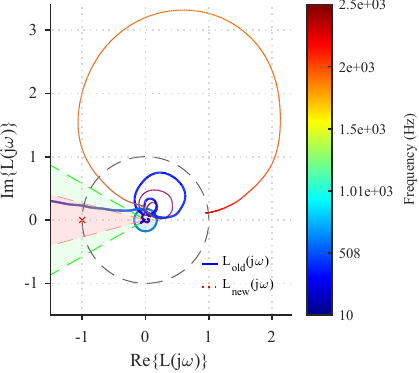}
    \caption{Case~1: Stability margins using frequency-domain model and negative-sequence impedance.}
    \label{fig:margins_harmonic_neg}
\end{figure}

Due to space constraints, only one case is presented for the Nyquist plot based on negative-sequence impedance. Fig. \ref{fig:margins_harmonic_neg} shows the stability margins obtained from the loop gain calculated using the negative-sequence impedance of the frequency-domain model for Case~1. The resulting conclusions are consistent with those drawn from the positive-sequence analysis. 

\begin{figure}[!t]
    \centering
    \includegraphics[width=1\linewidth]{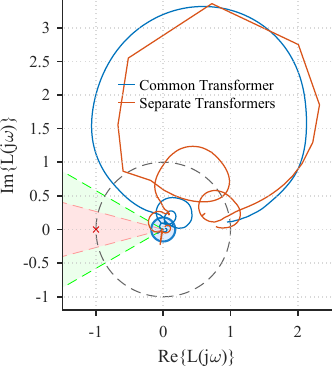}
    \caption{Comparison of stability margins for study cases~1 and~2 using the frequency-domain model.}
    \label{fig:Tx_compare}
\end{figure}

Introducing separate transformers in Case~2 shifts the resonance frequencies observed in Case~1 and introduces four additional resonance frequencies. Despite this, Case~2 exhibits improved stability margins, with a worst-case phase margin of $58.1^\circ$ compared to $42.4^\circ$ in Case~1, as shown in Fig.~\ref{fig:Tx_compare}. This improvement is attributed to the galvanic isolation and added damping provided by the two transformers, which increases the electrical distance between the OWPPs and the grid and thereby weakens their direct resonance interaction.

\subsection{Impedance specification method validation}

To validate the impedance-specification formulation in \eqref{eq:Zowpp_new_bound_wgc}, the phase margin is evaluated for all gain-crossover frequencies in the considered study cases. If the Nyquist locus intersects the unit circle within the \emph{critical area}, then there exists at least one frequency at which the phase margin falls below the minimum requirement. In that situation, the impedance of the newly connected plant (here, $\mathrm{OWPP}2$) exceeds the allowable limit given by \eqref{eq:Zowpp_new_bound_wgc}. As shown in Figs.~\ref{fig:margins_Thévenin_pos} and \ref{fig:margins_harmonic_pos}, the Nyquist locus of $L_{\mathrm{new}}$ does not intersect the unit circle within the critical area. Therefore, for all identified resonance (gain-crossover) frequencies, $\lvert Z_{\mathrm{OWPP}2}\rvert$ remains below the corresponding limit in both cases, as summarized in Table~\ref{tab:validation_case1}.

To validate that a newly connected plant with impedance exceeding the proposed limit violates the minimum phase-margin requirement, the impedance of $\mathrm{OWPP}2$ is emulated using a frequency-dependent network equivalent (FDNE) block. First, the frequency response of the FDNE block is verified against that of the original $\mathrm{OWPP}2$ model. The responses match closely, indicating that the FDNE can be used to selectively modify the impedance at the critical frequencies. After each impedance modification, the model is re-checked to ensure stable operation with the grid.

\begin{figure}
    \centering
    \includegraphics[width=1\linewidth]{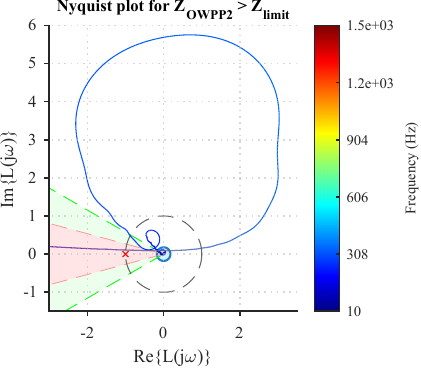}
    \caption{Case 1: Stability margins using a Thévenin-equivalent grid and positive-sequence impedance when $Z_{\mathrm{OWPP2}} > Z_{\mathrm{limit}}$.}
    \label{fig:margins_emt_greater}
\end{figure}

\begin{figure}
    \centering
    \includegraphics[width=1\linewidth]{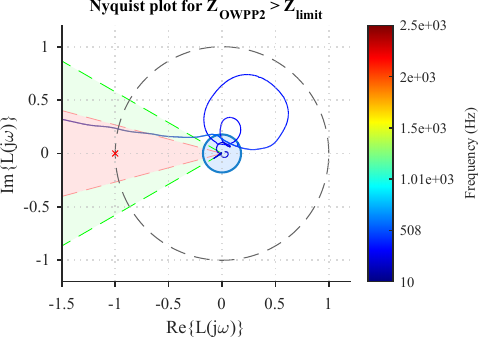}
    \caption{Case 1: Stability margins using frequency-domain model and positive-sequence impedance when $Z_{\mathrm{OWPP2}} > Z_{\mathrm{limit}}$.}
    \label{fig:margins_harmonic_greater}
\end{figure}

Increasing the impedance at selected critical frequencies results in $\lvert Z_{\mathrm{OWPP}2}\rvert$ exceeding the corresponding limit at certain gain-crossover frequencies. Consequently, the Nyquist locus intersects the unit circle within the critical area, thereby violating the minimum phase-margin requirement. This behavior is illustrated in Figs.~\ref{fig:margins_emt_greater} and \ref{fig:margins_harmonic_greater}, with the corresponding values of $\lvert Z_{\mathrm{OWPP}2}\rvert$ and $Z_{\mathrm{limit}}$ summarized in Table~\ref{tab:validation_case2}.

\begin{table}[!t]
\caption{Validation for Case studies in Figs.~\ref{fig:margins_Thévenin_pos}--\ref{fig:margins_harmonic_pos}, where $\lvert Z_{\mathrm{OWPP}2}\rvert < Z_{\mathrm{limit}}$.}
\label{tab:validation_case1}
\centering
\renewcommand{\arraystretch}{1.1}
\begin{tabular}{c c c c}
\hline
\textbf{Fig.} & \textbf{Crossover Freq. (Hz)} &
$\boldsymbol{\lvert Z_{\mathrm{OWPP}2}\rvert\;(\Omega)}$ &
$\boldsymbol{Z_{\mathrm{limit}}\;(\Omega)}$ \\
\hline
\ref{fig:margins_Thévenin_pos}  & 354.07  & 11.22 & 201.27 \\
\ref{fig:margins_Thévenin_pos}  & 561.60  &  1.98 &  22.03 \\
\ref{fig:margins_Thévenin_pos} & 997.49  & 24.85 & 718.59 \\
\ref{fig:margins_Thévenin_pos}  & 1370.00 &  6.77 & 101.72 \\
\hline
\ref{fig:margins_harmonic_pos} & 1324.5  &  7.88 &  72.0 \\
\ref{fig:margins_harmonic_pos} & 2414.6  &  2.09 &   168.71 \\
\hline

\end{tabular}
\end{table}

\subsection{Impact of Grid Model on Stability Assessment}
From Tables~\ref{tab:validation_case1} and~\ref{tab:validation_case2}, differences in the identified crossover frequencies can be observed. These differences are primarily attributed to the underlying modeling assumptions and impedance extraction methods. For instance, the Thévenin-equivalent grid model represents the offshore network using an impedance derived from the short-circuit level, whereas the frequency-domain model represents the full network using frequency-dependent characteristics derived from manufacturer data and field measurements, thereby capturing both passive and active impedance contributions. Although a Thévenin-equivalent representation may be sufficient when the objective is to assess internal resonance modes of an individual OWPP, the results illustrated in Figs.~\ref{fig:margins_Thévenin_pos} and~\ref{fig:margins_harmonic_pos} indicate that using a Thévenin grid for stability assessment of interconnected OWPPs can lead to inaccurate conclusions and unnecessary controller retuning for modes that may be damped by the actual network. 

Nonetheless, the results consistently show that, irrespective of the adopted grid model, violating the inequality in \eqref{eq:Zowpp_new_bound_wgc} causes the Nyquist locus to enter the critical region. These validations using OEM project models demonstrate that the proposed inequality provides a practical design criterion for parallel PPMs (or strings) to ensure that the existing plant remains compliant with the specified stability-margin requirements.

\begin{table}[!t]
\caption{Validation for Case studies in Figs.~\ref{fig:margins_emt_greater}--\ref{fig:margins_harmonic_greater}, where $\lvert Z_{\mathrm{OWPP}2}\rvert > Z_{\mathrm{limit}}$.}
\label{tab:validation_case2}
\centering
\renewcommand{\arraystretch}{1.1}
\begin{tabular}{c c c c}
\hline
\textbf{Fig.} & \textbf{Crossover Freq. (Hz)} &
$\boldsymbol{\lvert Z_{\mathrm{OWPP}2}\rvert\;(\Omega)}$ &
$\boldsymbol{Z_{\mathrm{limit}}\;(\Omega)}$ \\
\hline
\ref{fig:margins_emt_greater}  & 383.03  & 35.57 & 44.17 \\
\ref{fig:margins_emt_greater}  & 794.76  &  \textbf{70.03} &  \textbf{7.15} \\
\ref{fig:margins_emt_greater}  & 1075.5  & 75.03 & 107.76 \\
\hline
\ref{fig:margins_harmonic_greater} & 1628.8  &  \textbf{194.24} &  \textbf{156.76} \\
\hline

\end{tabular}
\end{table}

\section{Conclusion}
This paper has put forward a general impedance-based method for specifying the allowable impedance of a new power park module connected to a common connection point, such that the stability margins of an existing PPM are preserved. In addition, \emph{critical} and \emph{caution} regions are introduced on the Nyquist diagram, defined with respect to minimum phase- and gain-margin requirements, to complement the Nyquist stability criterion by enabling assessment of margin compliance and margin headroom.
The proposed design constraint is validated using study cases based on OEM EMT models for an actual OWPP project as well as frequency-domain models. The results show that integrating a parallel PPM can increase the stability margins of the existing plant. Moreover, enforcing the proposed impedance specification ensures that the existing PPM remains compliant and stable with the minimum margin requirements after the new connection is added. 

Overall, the validated approach provides a practical tool for planning and compliance studies of nearby separate PPMs in a cluster or strings inside a power plant, based on impedance data which can be obtained from frequency scans, frequency-domain models, or site measurements, thereby enabling developers and grid operators to assess stability margins and verify that minimum margin requirements are not violated.
In addition, since the method provides a direct criterion for determining whether a newly connected PPM satisfies the prescribed margin requirements at the point of connection, it can be readily integrated into grid-code compliance assessments and routine stability study practices to evaluate the impact of additional PPM integration.

\section*{Acknowledgments}
This work was supported by the ADOreD project of the European
Union’s Horizon Europe Research and Innovation Program under the Marie
Skłodowska-Curie Grant 101073554. The authors also thank Vattenfall for the opportunity to conduct this research and for their technical support and guidance throughout the project.

\bibliographystyle{IEEEtran}
\bibliography{References}

@ARTICLE{Mugambi2025,
   author = {Germano Rugendo Mugambi and Nicolae Darii and Hesam Khazraj and Oscar Saborío-Romano and Alin George Raducu and Ranjan Sharma and Nicolaos A. Cutululis},
   doi = {10.1016/j.rser.2025.115635},
   issn = {18790690},
   journal = {Renewable and Sustainable Energy Reviews},
   month = {7},
   publisher = {Elsevier Ltd},
   title = {Methodologies for offshore wind power plant dynamic stability analysis},
   volume = {216},
   year = {2025}
}

@ARTICLE{interaction,
  author={Lu, Minghui and Wang, Xiongfei and Loh, Poh Chiang and Blaabjerg, Frede},
  journal={IEEE Transactions on Power Electronics}, 
  title={Resonance Interaction of Multiparallel Grid-Connected Inverters With LCL Filter}, 
  year={2017},
  volume={32},
  number={2},
  pages={894-899},
  doi={10.1109/TPEL.2016.2585547}
  }

@inproceedings{LatinovicMilica2025,
   author = {Latinovi\'c Milica and Höhn Sebastian and Garcia Jose and Buchhagen Christoph and Sun Jian},
   booktitle = {Wind \& Solar Integration Workshop},
   month = {10},
   title = {ENHANCING SPECIFICATIONS ON HVDC DAMPING CAPABILITY AND IMPEDANCE MODELLING FOR SSO AND HARMONIC STABILITY STUDY},
   year = {2025}
}

@inproceedings{Buchhagen2016,
   author = {Christoph Buchhagen and Marlien Greve and Andreas Menze and Jochen Jung},
   city = {Vienna},
   booktitle = {15th International Workshop on Large-Scale Integration of Wind Power into Power Systems as well as Transmission Networks for Offshore Wind Farms},
   month = {11},
   title = {Harmonic Stability – Practical Experience of a {TSO}},
   year = {2016}
}

@article{Sun2011,
   author = {Jian Sun},
   doi = {10.1109/TPEL.2011.2136439},
   issn = {08858993},
   issue = {11},
   journal = {IEEE Transactions on Power Electronics},
   pages = {3075-3078},
   title = {Impedance-based stability criterion for grid-connected inverters},
   volume = {26},
   year = {2011}
}

@ARTICLE{Wang2020,
  author={Liao, Yicheng and Wang, Xiongfei},
  journal={IEEE Transactions on Power Electronics}, 
  title={Impedance-Based Stability Analysis for Interconnected Converter Systems With Open-Loop RHP Poles}, 
  year={2020},
  volume={35},
  number={4},
  pages={4388-4397},
  doi={10.1109/TPEL.2019.2939636}
  }

@inproceedings{Middlebrook,
   author = {R. D. Middlebrook},
   booktitle = {IEEE Ind. Appl. Soc. Annu. Meeting, pp. 91–107.},
   month = {11},
   title = {Input filter considerations in design and application of switching regulators},
   year = {1976}
}

@ARTICLE{Wildrick95,
  author={Wildrick, C.M. and Lee, F.C. and Cho, B.H. and Choi, B.},
  journal={IEEE Transactions on Power Electronics}, 
  title={A method of defining the load impedance specification for a stable distributed power system}, 
  year={1995},
  volume={10},
  number={3},
  pages={280-285},
  doi={10.1109/63.387992}
  }

@phdthesis{Wildrick93,
    author = {C. M. Wildrick},
    title = {Stability of distributed power supply systems},
    school = {Virginia Polytech. Inst. State Univ., Blacksburg, VA},
    month = {02},
    year = {1993}
}

@INPROCEEDINGS{Feng99,
  author={Xiaogang Feng and Zhihong Ye and Kun Xing and Lee, F.C. and Borojevic, D.},
  booktitle={APEC '99. Fourteenth Annual Applied Power Electronics Conference and Exposition. 1999 Conference Proceedings (Cat. No.99CH36285)}, 
  title={Individual load impedance specification for a stable DC distributed power system}, 
  year={1999},
  volume={2},
  number={},
  pages={923-929 vol.2},
  doi={10.1109/APEC.1999.750480}
  }

@ARTICLE{Feng202,
  author={Xiaogang Feng and Jinjun Liu and Lee, F.C.},
  journal={IEEE Transactions on Power Electronics}, 
  title={Impedance specifications for stable DC distributed power systems}, 
  year={2002},
  volume={17},
  number={2},
  pages={157-162},
  doi={10.1109/63.988825}
  }

@ARTICLE{Sun22,
  author={Sun, Jian},
  journal={IEEE Open Journal of Power Electronics}, 
  title={Frequency-Domain Stability Criteria for Converter-Based Power Systems}, 
  year={2022},
  volume={3},
  number={},
  pages={222-254},
  doi={10.1109/OJPEL.2022.3155568}
  }

@INPROCEEDINGS{Francisco24,
  author={Cifuentes Garcia, Francisco Javier and Roose, Thomas and Sakinci, {\"O}zg{\"u}r Can and Lee, Dongyeong and Dewangan, Lokesh and Avdiaj, Eros and Beerten, Jef},
  booktitle={2024 IEEE PES Innovative Smart Grid Technologies Europe (ISGT EUROPE)}, 
  title={Automated Frequency-Domain Small-Signal Stability Analysis of Electrical Energy Hubs}, 
  year={2024},
  volume={},
  number={},
  pages={1-6},
  doi={10.1109/ISGTEUROPE62998.2024.10863484}
  }

@ARTICLE{AminSSS,
  author={Amin, Mohammad and Molinas, Marta},
  journal={IEEE Transactions on Industry Applications}, 
  title={Small-Signal Stability Assessment of Power Electronics Based Power Systems: A Discussion of Impedance- and Eigenvalue-Based Methods}, 
  year={2017},
  volume={53},
  number={5},
  pages={5014-5030},
  doi={10.1109/TIA.2017.2712692}
  }

@ARTICLE{Cheah-Mane2023,
   author = {Marc Cheah-Mane and Agusti Egea-Alvarez and Eduardo Prieto-Araujo and Hasan Mehrjerdi and Oriol Gomis-Bellmunt and Lie Xu},
   doi = {10.1002/wene.453},
   issn = {2041840X},
   issue = {1},
   journal = {Wiley Interdisciplinary Reviews: Energy and Environment},
   month = {1},
   publisher = {John Wiley and Sons Ltd},
   title = {Modeling and analysis approaches for small-signal stability assessment of power-electronic-dominated systems},
   volume = {12},
   year = {2023}
}

@article{Z-toolFrancisco,
title = {Z-Tool: Frequency-domain characterization of EMT models for small-signal stability analysis},
journal = {Electric Power Systems Research},
volume = {252},
pages = {112405},
year = {2026},
issn = {0378-7796},
doi = {https://doi.org/10.1016/j.epsr.2025.112405},
url = {https://www.sciencedirect.com/science/article/pii/S0378779625009927},
author = {Francisco Javier {Cifuentes Garcia} and Jef Beerten}
}

@book{Maciejowski89,
    author = {J. M. Maciejowski},
    title = {Multivariable Feedback Design},
    publisher = {USA: Addison-Wesley} ,
    year = {1989}
}

@ARTICLE{ShahMIMO,
  author={Shah, Shahil and Parsa, Leila},
  journal={IEEE Transactions on Energy Conversion}, 
  title={Impedance Modeling of Three-Phase Voltage Source Converters in DQ, Sequence, and Phasor Domains}, 
  year={2017},
  volume={32},
  number={3},
  pages={1139-1150},
  doi={10.1109/TEC.2017.2698202}
  }

@ARTICLE{sunTwoPort_1,
  author={Sun, Jian},
  journal={IEEE Open Journal of Power Electronics}, 
  title={Two-Port Characterization and Transfer Immittances of AC-DC Converters—Part I: Modeling}, 
  year={2021},
  volume={2},
  number={},
  pages={440-462},
  doi={10.1109/OJPEL.2021.3104502}
  }

@techreport{AEMO_Report,
    author = {AEMO},
    title = {Analysis of Sub-synchronous Oscillations in West Murray Zone Power System in Australia},
    institution = {Australian Energy Market Operator},
    url = {https://www.aemo.com.au/initiatives/major-programs/engineering-roadmap/engineering-roadmap-execution-reports},
    year = {2025}
}

@ARTICLE{Sun2009,
  author={Sun, Jian},
  journal={IEEE Transactions on Power Electronics}, 
  title={Small-Signal Methods for AC Distributed Power Systems–A Review}, 
  year={2009},
  volume={24},
  number={11},
  pages={2545-2554},
  doi={10.1109/TPEL.2009.2029859}
  }

@misc{NREL,
title = {GIST (Grid Impedance Scan Tool) [SWR-22-73]},
author = {Shah, Shahil and Koralewicz, Prezmyslaw and Mendiola, Emanuel},
abstractNote = {NREL has developed GIST (Grid Impedance Scan Tool) to scan the impedance response of inverter-based resources such as land-based and offshore wind power plants, solar power plants, etc., from their electromagnetic transient simulation models to evaluate their impact on grid stability. GIST can scan impedance response of inverter-based resources from vendor-supplied black-box simulation models without requiring internal proprietary details, enabling detailed stability analysis of power systems with high levels of inverter-based generation.},
doi = {10.11578/dc.20221214.3},
url = {https://doi.org/10.11578/dc.20221214.3},
howpublished = {[Computer Software] \url{https://doi.org/10.11578/dc.20221214.3}},
year = {2022},
month = {sep}
}

@article{Nouri2021,
   author = {Behnam Nouri and Łukasz Kocewiak and Shahil Shah and Przemyslaw Koralewicz and Vahan Gevorgian and Poul Sørensen},
   doi = {10.1049/rpg2.12245},
   issn = {17521424},
   issue = {15},
   journal = {IET Renewable Power Generation},
   month = {11},
   pages = {3564-3576},
   publisher = {John Wiley and Sons Inc},
   title = {Test methodology for validation of multi-frequency models of renewable energy generators using small-signal perturbations},
   volume = {15},
   year = {2021}
}

@misc{Energinet2025,
   author = {Energinet},
   title = {Immittance Measurement ToolBox {(IMTB)}},
   url = {https://github.com/Energinet-SimTools/IMTB},
   year = {2025}
}

@article{Behnam_Nouri2025,
author = {Behnam Nouri  and Soroush Azarian  and José Zúñiga  and Matthias Dernbach },
title = {Challenges in small-signal stability compliance process for renewable energy-based power plants and a framework proposal for industrial application},
journal = {IET Conference Proceedings},
volume = {2025},
issue = {45},
pages = {263-267},
year = {2026},
doi = {10.1049/icp.2025.4311},

URL = {https://digital-library.theiet.org/doi/abs/10.1049/icp.2025.4311},
eprint = {https://digital-library.theiet.org/doi/pdf/10.1049/icp.2025.4311}
}

\balance
\begin{IEEEbiography}[{\includegraphics[width=1in,height=1.25in,clip,keepaspectratio]{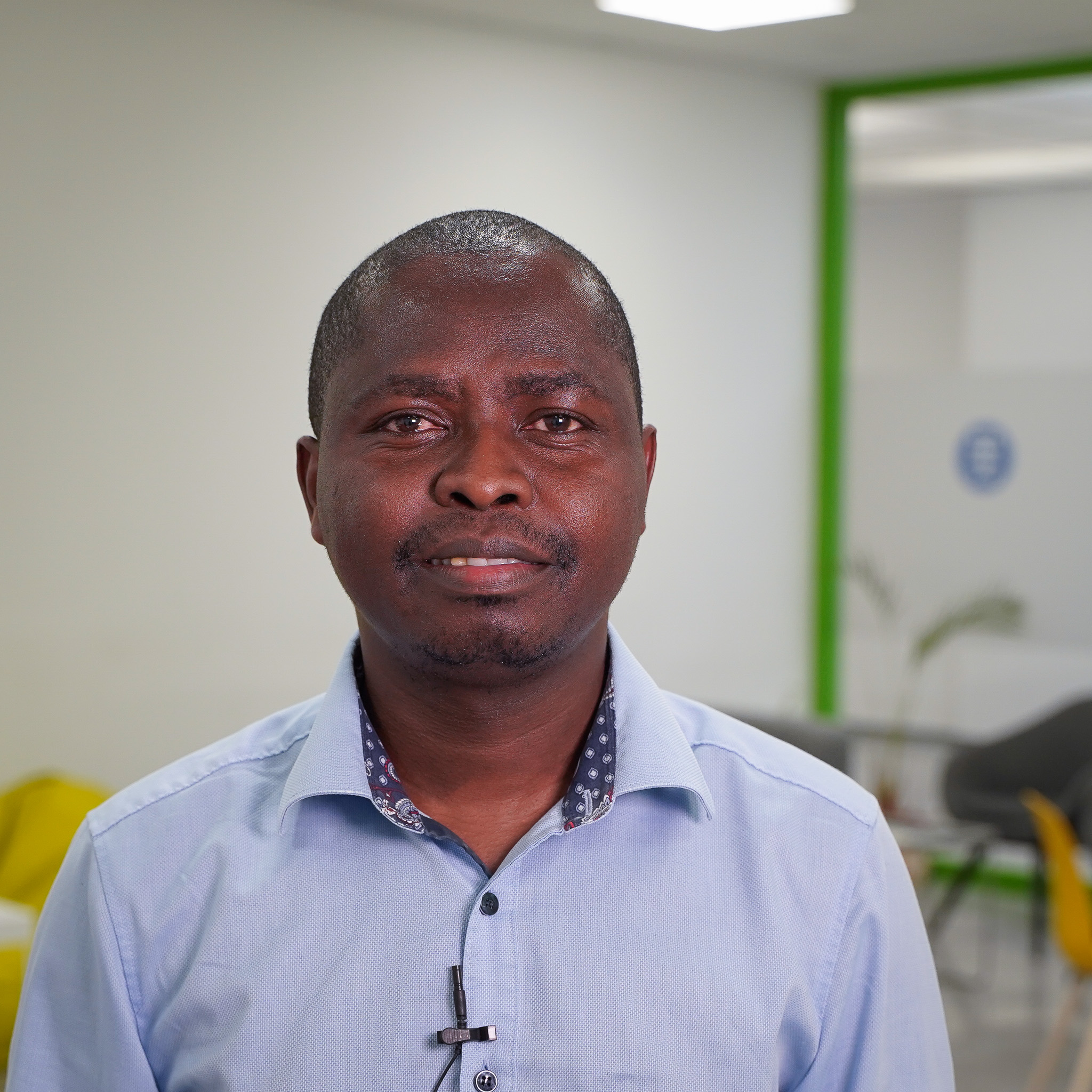}}]{Germano Mugambi} received the M.Sc. degree in Electrical Engineering from the University of Rostock, Germany, in 2018. He is currently working toward a Ph.D. degree with the Department of Wind and Energy Systems, Technical University of Denmark and Vattenfall Vindkraft A/S, Denmark.
His research interests include power system stability analysis, HVDC and integration and control of renewable energy systems.
\end{IEEEbiography}


\begin{IEEEbiography}[{\includegraphics[width=1in,height=1.25in,clip,keepaspectratio]{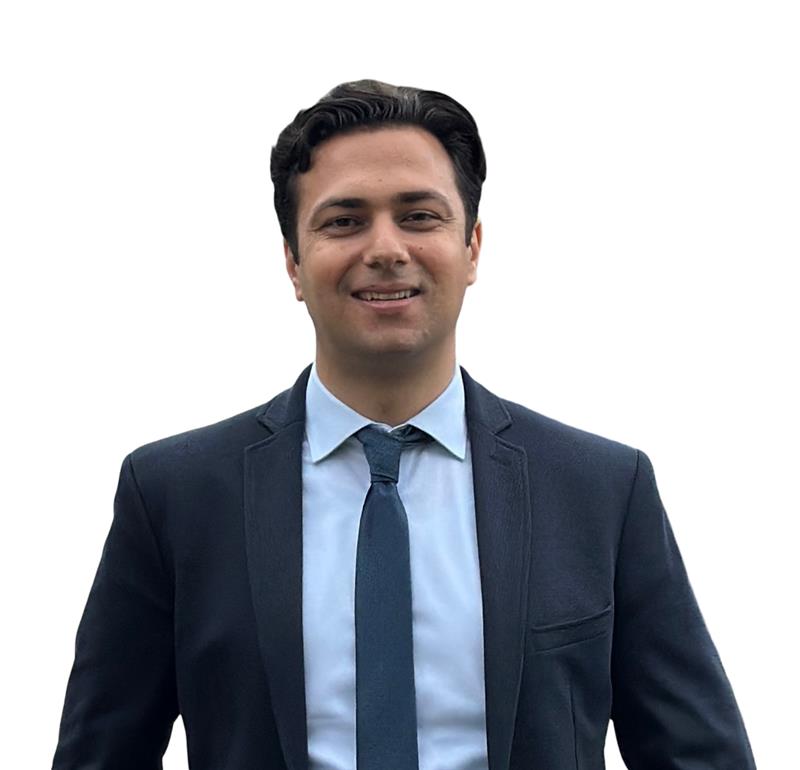}}]{Behnam Nouri} received the M.Sc. degree in power electronics and electrical machines from the University of Tehran, Tehran, Iran, in 2015. He joined the Department of Wind Energy at the Technical University of Denmark (DTU Wind Energy), Roskilde, Denmark, in 2018, where he received the Ph.D. degree in 2021.
From 2021 to 2023, he worked in the Power \&

\end{IEEEbiography}


\begin{IEEEbiography}[{\includegraphics%
	[width=1in,height=1.25in,clip,keepaspectratio]{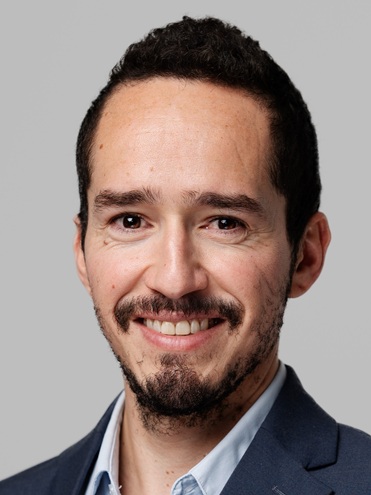}}]
	{Oscar Sabor\'{i}o-Romano} (Senior~Member, IEEE) received the B.Sc. (Hons.) degree in 
	electrical engineering from the University of Costa Rica, Costa Rica, in 2013. 
	In 2015, he received the M.Sc. degrees in electrical engineering and wind energy from Delft University of Technology, The Netherlands, and the Norwegian University of Science and Technology, Norway, respectively.
	He joined the Department of Wind and Energy Systems at the Technical University of 
	Denmark, Denmark, in 2016, where he received his PhD degree and is currently a Senior Advisor.
	His research interests include grid integration of renewables, HVDC, and
	modeling, control, and stability assessment of converters and converter-rich power systems.
\end{IEEEbiography}


\begin{IEEEbiography}[{\includegraphics[width=1in,height=1.25in,clip,keepaspectratio]{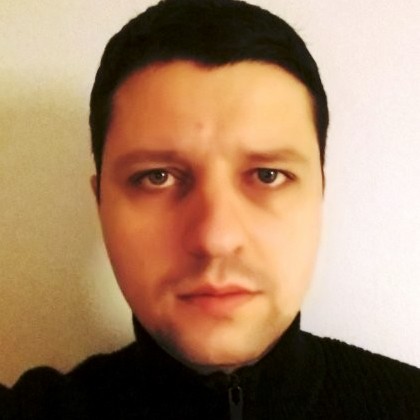}}]{George Alin Raducu} received the M.Sc. degree in Power Electronics and Drives from the Aalborg University, Denmark in 2007. With heavy background in Control Engineering, he has been working across the Renewable Energy sector as well as Power Electronics manufacturers over the last 20 years. Over the last 10 years he is affiliated with Vattenfall Vindkraft A/S holding the role of Head of Power Plant Control Department.
His research interests include Control Engineering, Renewable Energy integration, Power Plant Control and power system stability. 
\end{IEEEbiography}


\begin{IEEEbiography}[{\includegraphics[width=1in,height=1.25in,clip,keepaspectratio]{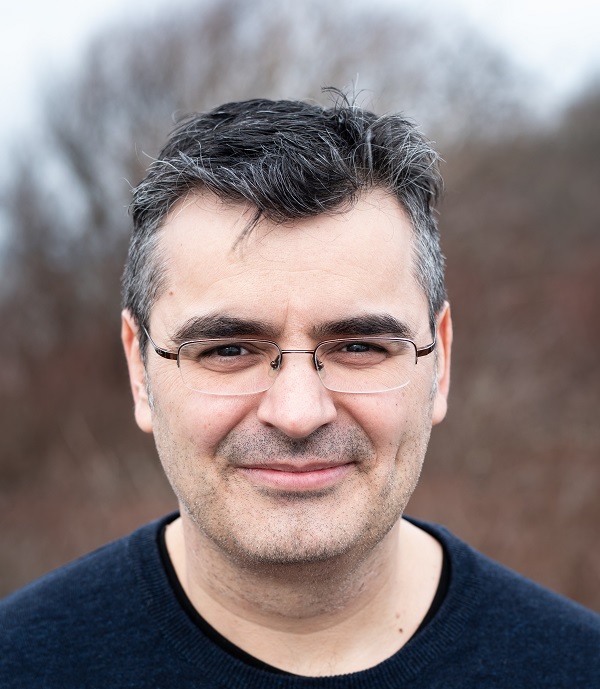}}]{Nicolaos A. Cutululis} (SM’ 16, M ’06) is a Professor in the Department of Wind and Energy Systems at the Technical University of Denmark. He holds a M.Sc. (1998) and a Ph.D (2005) in Automatic
Control. His main research area is operation and integration of wind power moving towards fully RES power systems, with a special focus on offshore
wind and HVDC. 
\end{IEEEbiography}

\vfill

\end{document}